\documentstyle[twoside,fleqn,espcrc2]{article}
\input epsf

\newcommand{\AmS}{{\protect\the\textfont2
  A\kern-.1667em\lower.5ex\hbox{M}\kern-.125emS}}

\title{Landau-Ginsberg Theory of Quark Confinement}

\author{Michael C. Ogilivie and Peter N. Meisinger\address{Department of Physics,
        Washington University, \\ 
        St. Louis, MO 63130 USA}%
        \thanks{We thank the U.S. Department of Energy
	for financial support under
	grant number DE-FG02-91-ER40628.}
}
       
\begin{document}

\begin{abstract}

We describe the SU(3) deconfinement transition using Landau-Ginsberg theory.
Drawing on perturbation theory and symmetry principles, we construct
the free energy as a function of temperature and the Polyakov loop. Once the
two adjustable parameters of the model are fixed, the pressure $p$, energy $%
\varepsilon $ and Polyakov loop expectation value $P_{F}$ are calculable
functions of temperature. An excellent fit to the continuum extrapolation of
lattice thermodynamics data can be achieved. In an extended form of the
model, the glueball potential is responsible for breaking scale invariance
at low temperatures. Three parameters are required, but the glueball mass
and the gluon condensate are calculable functions of temperature, along with
$p$, $\varepsilon $ and $P_{F}$.

\end{abstract}

\maketitle

\section{Theory}

We take \vspace{1pt}the free energy density $f\,$\ of the gluon plasma\ to
be a function of the temperature $T\,$and the fundamental representation
Polyakov loop $P$. The theory also depends on a renormalization group
invariant scale-setting parameter $\Lambda $. Perturbation theory gives a
free energy $f$ of the form $T^{4}f_{4}(P,g(T/\Lambda ))$.
Perturbation theory does not describe $f$
near the deconfining transition, but is probably adequate for
$T$ much greater than the deconfinement temperature
$T_d$\cite{AndersenBraaten99}\cite{BlaizotIancu99}.
Subleading terms, of the
form $T^{4-r}\Lambda ^{r}f_{r}(P,g(T/\Lambda ))$ , are likely needed to
describe the deconfining transition. Such terms are inherently
non-perturbative, due to the appearance of the factor $\Lambda ^{r}$. It
is easy to show that $\Delta \equiv \varepsilon -3p\,$is given by
\[
\Delta =\left[ 4-T\partial_T \right] \,f
\]
and therefore contains information about the subleading terms. Note that
$\Delta $ is also directly related to the finite temperature
contribution to the stress-energy tensor anomaly,
which depends in a non-trivial way on the Polyakov
loop\cite{MeisingerOgilvie95}.

\vspace{1pt}Given the close connection between $\Delta $ and the 
subleading terms in
$f$ which drive the deconfinement transition,
it is natural to examine \ the behavior of $\Delta (T)$ near $T_{d}$ as
measured in simulations. Using the data of Boyd et al for $SU(3)\,$%
lattice gauge theory\cite{BoydEngels96},
we find that $\Delta (T) \propto T^{2}$ over a
large range of temperatures above $T_{d}$.
This suggests that a term in $f$
proportional to $T^{2}$ plays an important role in the deconfinement
transition. It is also necessary to have a term
proportional to $T^{0}$ (independent of $T\,$), so that there is a non-zero
free energy density difference between confined and deconfined phases at
very low temperatures. Thus we conjecture the simple form for the free
energy
\[
f(T,P)=T^{4}f_{4}(P)+T^{2}\Lambda ^{2}f_{2}(P)+\Lambda ^{4}f_{0}(P)
\]
where $f_{0}$ must favor the confined phase
to yield confinement at arbitrarily low
temperatures.

We look at the one loop perturbative result for guidance on the possible
forms for $f_{r}$. \vspace{1pt}We define the eigenvalues $q_{j}$ by
diagonalizing the fundamental representation Polyakov loop $P\,$:
$P_{jk} =  exp \left[ i \pi q_{j} \right] \delta _{jk} $.
\vspace{1pt}The free energy for gluons in a constant $A_{0}$ background is
\cite{GrossPisarski81}\cite{Weiss8182}:
\begin{eqnarray*}
f_{g}(q) &=&\frac{2}{\beta } Tr_A \int
\frac{d^{3}k}{(2\pi )^{3}}\ln \left[ 1-e^{-\beta \mathbf{\omega }_{k}} P
\right]  \\
&=&-\frac{2}{\beta }Tr_A \int \frac{%
d^{3}k}{(2\pi )^{3}}\sum_{n=1}^{\infty }\frac{1}{n}e^{-n\beta \mathbf{\omega
}_{k}}P^n \\
&=&\frac{2\pi ^{2}T^{4}}{3}\sum_{j,k=1}^{N}(1-\frac{1}{N}\delta _{jk})
B_{4}\left( \frac{\left| \Delta q_{jk}\right| _{2}}{2}\right)
\end{eqnarray*}
where $\left| \Delta q_{jk}\right| _{2}\equiv
\left( q_j - q_k \right) mod(2)
$ and $B_4$ is the fourth Bernoulli polynomial, given by
$B_{4}(x)=x^{4}-2x^{3}+x^{2}-\frac{1}{30}\,$.
The free energy is a sum of terms, each of which represents
field configurations in which a net number of $n$ gluons go around
space-time in the Euclidean time direction.

\vspace{-0.33in}
\begin{figure}[htb]
\epsfxsize=75mm \epsfbox{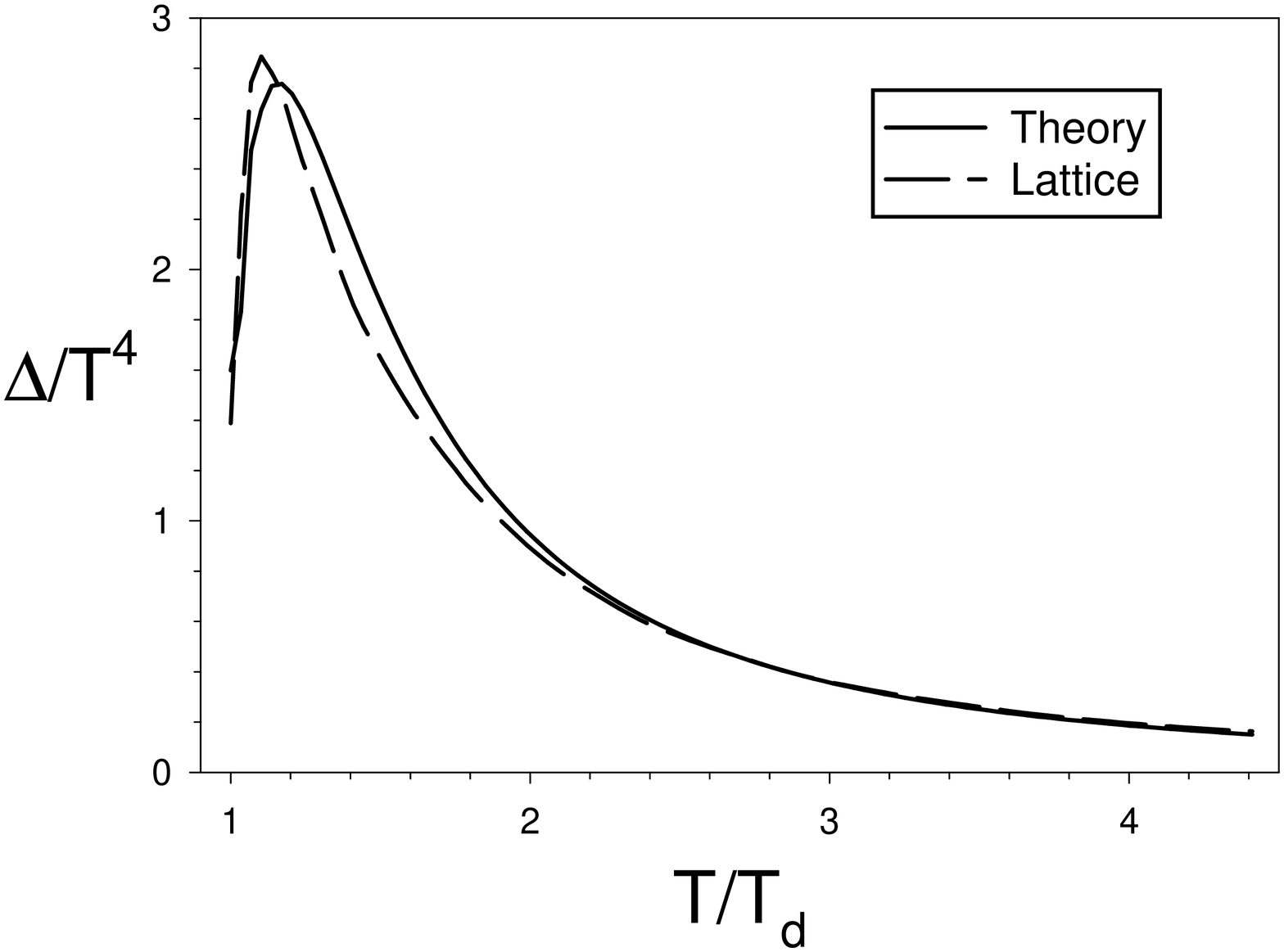}
\vspace{-0.33in}
\caption{$\Delta / T^4$ versus $T$.}
\label{fig:fig1}
\end{figure}
\vspace{-0.33in}

\section{Simple Model}

In imitation of perturbation
theory, we use the Bernoulli polynomial to construct  $f_{0}$, $f_{2}$ and $%
f_{4}$ as polynomials in the $q\,$variables with the appropriate symmetries.
There are two
inequivalent directions in the Cartan subalgebra of $SU(3)$, $%
\lambda _{3}$ and $\lambda _{8}$. Confinement is\ achieved by motion in the $%
\lambda _{3}$ direction away from $A_{0}=0$. 
Defining $P_F = Tr_F (P)$, we parametrize motion along the
line $Im P_{F}=0$ by

\[
P_{F}\,(\psi )=1+2\cos \left[ 2\pi \left( 1-\psi \right) / 3 \right]
\]
with $\psi =0$ giving $P_{F}=0$, and $\psi =1$ corresponding to $A_{0}=0$
and $P_{F}=3$. We take the free energy density to have the form
\[
f(\psi ,T)=a T^{4}\left( \psi ^{4}-\frac{2}{3}\psi
^{3}+\psi ^{2}\right) +\left( b+cT^{2}\right) \psi ^{2}
\]
where $a = 4 \pi^2 / 15 $; $b$ and $c$ fix the critical properties.
This potential can be extended to the entire Lie algebra, and contains all
required symmetries. 
For low temperatures, the $b\psi
^{2}$ term dominates. If $b>0$, the system will be confined. The parameter $%
b\,$can be interpreted as the free energy difference at $T=0$ between the $%
\psi =0$ confined phase and the fully deconfined $\psi =1$ phase.

\section{Results}

\vspace{1pt}The above potential has built in the correct low and high
temperature behavior, and has two free parameters, $b\,$and $c$. We can use
one of these to set the overall scale by fixing the deconfinement
temperature. To determine the remaining parameter,
we fit the lattice data for $\Delta $ at $N_{t}=8$, which is well measured
and a good approximation to the continuum limit\cite{BoydEngels96}.
With $T_d = 0.272\, GeV$, we obtain $b^{1/4} = 0.356\, GeV$
and $c^{1/2} = 0.313\, GeV$.
The results of our fitting procedure are shown in figures 1, 2 and 3 for $%
\Delta $, $p$ and $\varepsilon $. The agreement is good throughout the range
$T_{d}-4T_{d}$. The discrepancy in the high-temperature behavior
of $p$ and $\varepsilon$ is probably accounted for by HTL-improved
perturbation theory\cite{AndersenBraaten99}\cite{BlaizotIancu99}.

\vspace{-0.33in}
\begin{figure}[htb]
\epsfxsize=75mm \epsfbox{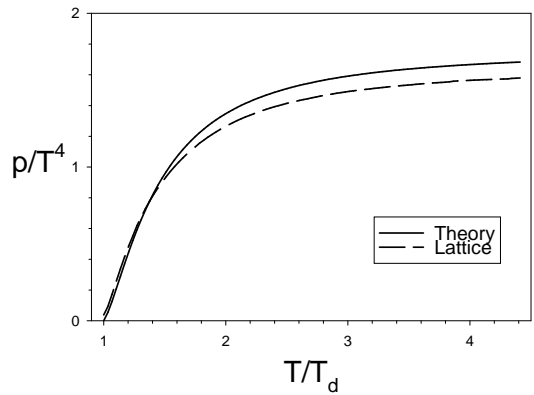}
\vspace{-0.33in}
\caption{$p / T^4$ versus $T$.}
\label{fig:fig2}
\end{figure}
\vspace{-0.33in}

\section{Extended Model}

The physical origin of the
parameters $b\,$and $c$ above is obscure.
Since the trace of the stress-energy tensor $%
\theta _{\mu }^{\mu }\,$couples to the scalar glueball, we
introduce a scalar glueball field $\phi $ as the source of scale symmetry
breaking in an extended model. For $SU(3)$, our extended model is
\begin{eqnarray*}
f&=&a T^{4}\left( \psi ^{4}-\frac{2}{3}+\psi
^{2}\right)
+\left( \alpha \phi ^{4}+\beta \phi ^{2}T^{2}\right) \psi^{2} \\
&\phantom{=}&+\lambda \phi ^{4}\log \left( \frac{\phi ^{2}}{e^{1/2}\mu ^{2}}\right)
.
\end{eqnarray*}

Spontaneous symmetry breaking of $\phi $ via a Coleman-Weinberg
potential introduces the scale $\mu $.
If we make the identification $\phi ^{4}\propto
Tr\,\left( F_{\mu \nu }^{2}\right) $, the $T=0$ potential for $\phi $ can be
derived in a variety of ways: 1) renormalization
group \cite{MatinianSavvidy78}; 2) explicit calculation for constant
fields\cite{NielsenOlesen78};
3) stress-energy tensor anomaly \cite{PagelsTomboulis78};
4) stress-energy sum rules \cite{CornwallSoni84}. The values of $%
\lambda $ and $\mu $ can be determined from the values of the gluon
condensate and the glueball mass. 
The $T = 0$ condensate from the
Coleman-Weinberg potential is given by $-2\lambda \phi
^{4}\rightarrow -2\lambda \mu ^{4}$and the glueball mass is given by $%
M_{s}^{2}=8\lambda \mu ^{2}$. A similar glueball potential has been
used to model the chiral transition \cite{CampbellEllis90}.
A coupling between $\phi$
and $P\ $ can be inferred from perturbation theory \cite{MeisingerOgilvie95}
\cite{MeisingerOgilvie97};
similar
couplings to the chiral order parameter exist \cite{MeisingerOgilvie96}
\cite{ChandrasekharanHuang96}\cite{Stephanov96}.

\vspace{-0.33in}
\begin{figure}[htb]
\epsfxsize=75mm \epsfbox{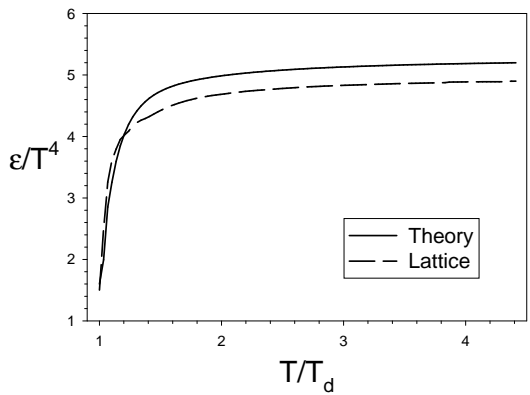}
\vspace{-0.33in}
\caption{$\varepsilon / T^4$ versus $T$.}
\label{fig:fig3}
\end{figure}
\vspace{-0.33in}

We have found values for the parameters $\alpha $, $\beta $, $%
\lambda \,$and $\mu \,$which mimic the behavior of our simpler model near $%
T_{d}$. Our extended model has a potentially fatal problem associated with
the restoration of scale symmetry. For plausible values of the
gluon condensate and glueball mass, restoration of scale symmetry at $T_{d}$
leads to a single abrupt phase transition incompatible with lattice data.
The alternative, with unrealistic values, is restoration above $T_{d}$
via a first order transition, which would be observable in lattice data.
This argues against
any simple role of the glueball in the thermodynamics of the gluon plasma.

\end{document}